\documentclass[12pt,preprint]{aastex}

\shorttitle{Cluster Core Binary Fractions}
\shortauthors{Hurley, Aarseth \& Shara}

\slugcomment{To appear in the Astrophysical Journal}

\begin{document}

\title{The core binary fractions of star clusters from realistic simulations}

\author{Jarrod R. Hurley}
\affil{Centre for Astrophysics and Supercomputing, 
        Swinburne University of Technology, \\
        P.O. Box 218, VIC 3122, Australia}
\email{jhurley@swin.edu.au}

\author{Sverre J. Aarseth}
\affil{Institute of Astronomy, 
        Madingley Road, 
        Cambridge CB3 0HA, UK}
\email{sverre@ast.cam.ac.uk}

\author{Michael M. Shara}
\affil{Department of Astrophysics,
        American Museum of Natural History, \\
        Central Park West at 79th Street, 
        New York, NY 10024}
\email{mshara@amnh.org}

\begin{abstract}
We investigate the evolution of binary fractions in star clusters using $N$-body 
models of up to $100\,000$ stars. 
Primordial binary frequencies in these models range from 5\% to 50\%. 
Simulations are performed with the {\tt NBODY4} code and include 
a full mass spectrum of stars, stellar evolution, binary evolution and the tidal 
field of the Galaxy. 

We find that the overall 
binary fraction of a cluster almost always remains close to the primordial value, 
except at late times when a cluster is near dissolution. 
A critical exception occurs in the central regions where we observe a marked 
{\it increase} in binary fraction with time 
-- a simulation starting with $100\,000$ stars and 5\% binaries reached a core binary 
frequency as high as 40\% at the end of the core-collapse phase (occurring at $16\,$Gyr with 
$\sim 20\,000$ stars remaining). 
Binaries are destroyed in the core by a variety of processes as a cluster evolves, 
but the combination of mass-segregation and creation of new binaries in exchange 
interactions produces the observed increase in relative number. 
We also find that binaries are cycled into and out of cluster cores in a manner that is 
analogous to convection in stars. 
For models of $100\,000$ stars we show that the evolution of the core-radius 
up to the end of the initial phase of core-collapse is not affected by the exact value of the 
primordial binary frequency (for frequencies of 10\% or less). 
We discuss the ramifications of our results for the likely primordial binary content 
of globular clusters. 
\end{abstract}


\section{Introduction}
\label{s:intro}

The binary content of a globular cluster is important in determining the frequency and nature 
of cluster stellar exotica, as well as the dynamical evolution of the cluster. 
It has long been recognized that binary formation is inevitable in a 
self-gravitating system\footnote{The ten-body gravitational calculations 
of von Hoerner (1960) are the earliest $N$-body calculations published.  
They were continued until the first binary formed, 
at which point the calculations were halted.}. 
Indeed, the presence of binaries as a central energy source is vital 
to avoid complete core-collapse (Goodman \& Hut 1989). 
However, only more recently has it been realised that globular clusters 
must also have formed with a sizeable binary population 
(see Hut et al. 1992 for an early review). 
That globular clusters harbour a mixture of dynamically formed and primordial 
binaries can be used to understand observations of their stellar 
content, such as the diverse blue straggler population in 47 Tucanae 
(Mapelli et al. 2004). 

Knowledge of the likely primordial binary fraction of globular clusters is 
essential as input to models of globular cluster evolution. 
It also provides a constraint on the cluster formation process. 
Considering that the presence of binaries in the cluster core 
has a pronounced effect on the core properties 
and cluster evolution (Hut 1996), knowledge of the central binary frequency is also 
important. 
Indications are that this is relatively small
 -- of the order of 20\% (e.g. Bellazzini et al. 2002) or less (e.g. Cool \& Bolton 2002) -- 
when compared to the frequencies of binaries observed in the solar 
neighbourhood (Duquennoy \& Mayor 1991) 
and open clusters such as M67 (Fan et al. 1996) 
which are of the order of 50\%.  

It would be particularly useful to take measurements of the 
current binary fraction in globular clusters 
-- whether that be in the core or outer regions -- 
and extrapolate backwards to gain a reliable determination of the 
primordial binary content. 
However, processes involved in the intervening cluster evolution 
make this difficult. 
For example, binaries can be formed and destroyed in a variety of interactions 
between cluster members (Hurley \& Shara 2002). 
Binaries will on average be more massive than single stars and thus are 
affected differently by mass segregation. 
Also, the escape rates of single stars and binaries will differ. 
Finally, the internal evolution of the components of binaries can also lead 
to binaries' destruction. 

Current simulation techniques have been designed to model these (and other) 
processes (Aarseth 2003) and have reached the level of sophistication 
required to produce realistic cluster models. 
In this way the link between primordial and current cluster binary populations 
can be investigated directly (e.g. Hurley et al. 2005; Ivanova et al. 2005). 
Aarseth (1996) conducted an $N$-body simulation starting with $10\,000$ stars 
and a 5\% binary frequency where notably the stars were drawn from a 
realistic initial mass function (IMF), the cluster was subject to the tidal 
field of the Galaxy, and both stellar and binary evolution were modelled. 
This model cluster had a half-life of about $2\,$Gyr at which point the 
core binary frequency had risen to 20\% primarily owing to mass-segregation. 
Thus binaries were not preferentially depleted. 
In this case it was not necessary to include a large initial binary fraction 
in order to halt core-collapse and yield a significant observed abundance 
in the central regions. 
The earlier work of McMillan \& Hut (1994) reported $N$-body simulations 
of $2\,000$ stars or less and binary frequencies in the range of 5-20\%. 
They included the Galactic tidal field but only considered point-mass dynamics. 
McMillan \& Hut (1994) showed that there is a critical primordial binary frequency 
of 10-15\% below which the binaries are destroyed before the cluster dissolves 
owing to the tidal field. 
Furthermore, they found that above this critical value there exists a minimum possible 
binary mass fraction for the cluster -- this result could be used with observations of 
present-day binary frequency to place limits on the primordial frequency. 
We note that the McMillan \& Hut (1994) simulations were restricted 
to equal-mass stars, and the binaries were a factor of two heavier then single stars 
-- this could give misleading results when applied to real clusters\footnote{
Binaries would naturally be twice as massive as single stars on average if 
binaries form by random pairings independent of the stellar IMF. 
In general correlated masses are assumed (e.g. Kroupa 1995) although 
the exact situation is unclear 
-- the recent survey of stars in the solar neighborhood and in young open clusters 
compiled by Halbwachs et al. (2003) shows a distribution of mass-ratios, $q$, 
with a broad peak between $0.2 - 0.7$ but also a sharp peak for $q > 0.8$. 
}.

These $N$-body simulations were definitely in the open cluster regime. 
Dynamical processes which destroy (and equally may create) cluster binaries 
are density dependent. 
In addition, the central stellar density of a cluster is a function of the number, $N$, 
of cluster members. 
Thus, it is not clear that these prior results apply to globular cluster conditions. 
More recently Ivanova et al. (2005) have conducted Monte Carlo simulations 
of clusters with up to $5 \times 10^5$ members and core number densities 
ranging from $10^3 - 10^6 \, {\rm stars} \, {\rm pc}^{-3}$. 
They show that an initial binary frequency of 100\% is required to produce a 
current core binary frequency of 10\% for a globular cluster such as 47 Tucanae. 
Depletion of binaries in the cluster core is found to be the result of stellar 
evolution processes as well as three- and four-body dynamical interactions. 
It is our intention in this paper to test these claims by utilising direct $N$-body 
simulations of star clusters with up to $N = 100\,000$ members initially. 

One aspect that will affect the evolution of the cluster binary population is 
the orbital parameters of the primordial binaries -- in particular the initial 
ratio of {\it hard} to {\it soft} binaries. 
The boundary between these two regimes is determined by the mean 
kinetic energy of the cluster stars (with binaries represented by their 
center-of-mass motion) 
where hard binaries have a binding energy 
in excess of 2/3 of the mean kinetic energy (Hut et al. 1992).
We note that a useful estimate for the boundary in terms of the binary orbital separation 
is given by twice the cluster half-mass radius divided by $N$.
In three-body single--binary star interactions hard binaries tend to harden 
and provide kinetic heating for the cluster (Heggie 1975; Hut 1983). 
Soft binaries are less strongly bound (and thus on average are wider) 
and are efficiently destroyed in three- and four-body encounters. 
As noted by Hut et al. (1992) it is for this reason that soft binaries 
are not generally included in cluster models. 
A common misconception is that the omission of soft binaries is to aid 
the speed of simulation; however it is binaries near the hard/soft 
boundary that provide the main threat to efficient simulation (Aarseth 2003). 
The omission is more a realisation that soft binaries have little impact on 
the cluster dynamics or exotic star formation and so the focus is on the 
more {\it meaningful} binaries, so to speak. 
Neglecting soft binaries has the capacity to alter binary fractions in the 
halo of a model cluster as binary encounters tend to occur in or near the 
cluster core. 
For this reason we will attempt to account for any omitted soft binary 
populations when making binary fraction comparisons. 

Our simulation method and initial conditions are detailed in Sections 2 and 3. 
Results are given in Section 4 followed by discussion in Section 5. 
We briefly summarize our results in Section 6.

\section{Models}
\label{s:models}

All simulations utilized in this work were performed using 
the {\tt NBODY4} code (Aarseth 1999) 
on GRAPE-6 boards (Makino 2002) 
located at the American Museum of Natural History. 
{\tt NBODY4} uses the 4th-order Hermite integration scheme 
and an individual timestep algorithm to follow the orbits of cluster 
members and invokes regularization schemes to deal with the 
internal evolution of small-$N$ subsystems 
(see Aarseth 2003 for details). 
Stellar and binary evolution of the cluster stars are performed in 
concert with the dynamical integration as described in Hurley et al. (2001).  

\begin{figure}
\plotone{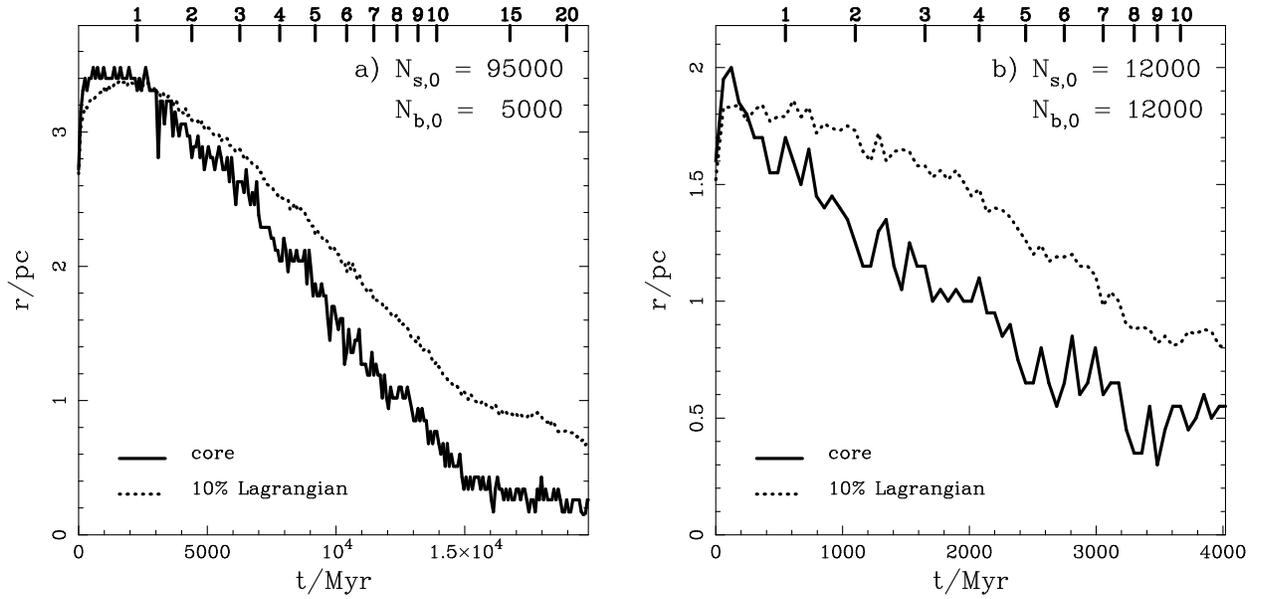}
\caption{
Evolution of the core radius (solid line) and the radius containing 
the inner 10\% of the cluster mass (dotted line) for: 
a) the K100-5 simulation; and 
b) the K24-50 simulation.  
The numbers across the top show the number of half-mass 
relaxation times that have elapsed. 
Note that $N_{{\rm s},0}$ and $N_{{\rm b},0}$ refer to the number of single 
stars and binaries, respectively, in the starting model. 
\label{f:fig1}}
\end{figure}

The results of four extensive simulations (detailed below) form the 
dataset for this paper. 
We will make use of data from two simulations that have previously 
been reported in the literature 
-- a simulation starting with $95\,000$ single stars and $5\,000$ binaries 
(Shara \& Hurley 2006)  and a simulation starting with $12\,000$ single stars and 
$12\,000$ binaries (Hurley et al. 2005). 
The former contained $100\,000$ members at birth, if we count each 
binary as one object, and thus had a primordial binary frequency of 5\%. 
We will refer to this as the K100-5 simulation. 
After about nine Gyr of evolution the cluster membership was reduced 
by half and at an age of $15 - 16\,$Gyr the model cluster had reached the end of 
the main core-collapse phase 
(associated with a minimum in core-radius, after which the size of the core stabilizes, 
in relative terms). 
Figure~\ref{f:fig1}a shows the behaviour of the core radius as the K100-5  
model evolves. 
Also shown is the 10\% Lagrangian radius -- the radius which encloses the 
inner 10\% of the cluster by mass. 
From Figure~\ref{f:fig1}a we see that initially the inner regions of the cluster expand 
owing to stellar evolution mass-loss before two-body effects take over and drive a 
prolonged period of contraction. 
When the cluster is about 12 half-mass relaxation times old (as denoted across the top 
of Fig.~\ref{f:fig1}a) the core radius reaches a minimum of $0.17\,$pc 
and the main core-collapse phase is halted. 
The 10\% Lagrangian radius at this point is $0.94\,$pc. 
The core density of the model begins at $10^2 \, {\rm stars} \, {\rm pc}^{-3}$ and 
increases to a maximum of $10^4 \, {\rm stars} \, {\rm pc}^{-3}$ just before 
termination of the model at $20\,$Gyr. 

The core radius in Figure~\ref{f:fig1} is actually the density radius commonly 
used in $N$-body simulations (Casertano \& Hut 1985). 
It is calculated from the density weighted average of the distance of each 
star from the density centre (Aarseth 2003). 
This definition, in combination with the effects of three-body interactions 
and the movement of binaries across the core boundary,  
allows for the small-scale fluctuations in core radius observed in Figure~\ref{f:fig1}. 
Such fluctuations could be smoothed out (see Heggie, Hut \& Trenti 2006, 
for example) but we have chosen not to do this. 
This $N$-body core-radius is distinct from observational determinations of 
core-radius calculated, for example, from the surface brightness profile (SBP) of a cluster. 
As discussed by Wilkinson et al. (2003) there is no general relation between the two 
quantities but usually the $N$-body value is the lesser of the two. 
This is supported by an in-depth analysis of the core-radius evolution of the 
K100-5 simulation that will be presented in an upcoming paper (Hurley, 2007, 
in preparation). 
Preliminary results show that the the core-radius obtained from the two-dimensional 
projected SBP of the K100-5 model agrees well with the $N$-body core radius for the 
first $7\,$Gyr of evolution but is about twice as large by the time the model reaches 
$16\,$Gyr of age. 
Thus the binary fraction within the 10\% Lagrangian radius may often be a 
better number to compare with central binary fractions quoted for real 
clusters and we will give both this and the core binary fraction in our results. 

The second model had a primordial binary frequency of 50\% and was 
tailored to investigate the evolution and stellar populations of the old open 
cluster M67. 
It had $24\,000$ members at birth and we will refer to this as the K24-50 simulation.  
It had a half-life of about $2\,$Gyr and after $4\,$Gyr of evolution only 
$2\,000$ stars and binaries remained. 
The core density was about $10^2 \, {\rm stars} \, {\rm pc}^{-3}$ on average,  
reaching a maximum of $350 \, {\rm stars} \, {\rm pc}^{-3}$ at $3\,480\,$Myr 
with a corresponding core radius of $0.3\,$pc. 
Figure~\ref{f:fig1}b shows the evolution of the core and 
10\% Lagrangian radii for the K24-50 simulation. 

To investigate the evolution of binary fractions across a range of 
star cluster models we will also make use of two simulations that 
have yet to be published. 
These are a simulation that started with $90\,000$ single stars and $10\,000$ 
binaries (K100-10) and a simulation that started with $40\,000$ single stars 
and $10\,000$ binaries (K50-20). 
In Table~\ref{t:table1} we summarize the properties of the four simulations. 

For each model the initial setup is as follows. 
Masses for the single stars are drawn from the IMF  
of Kroupa, Tout \& Gilmore (1993) between the mass limits of 0.1 and $50 M_\odot$.
Each binary mass is chosen from the IMF of Kroupa, Tout \& Gilmore (1991), 
as this had not been corrected for the effect of binaries, 
and the component masses are set by choosing a mass-ratio from a uniform distribution. 
We assume that all stars are on the zero-age main sequence (ZAMS) when 
the simulation begins and that any residual gas from the star formation 
process has been removed. 
We use a Plummer density profile (Aarseth, H\'{e}non \& Wielen 1974) and assume the stars 
and binaries are in virial equilibrium when assigning the initial positions and 
velocities. 
There is no primordial segregation by mass, binary properties, or any other 
discriminating factor in these models. 
Each cluster is subject to a standard Galactic tidal field 
-- a circular orbit in the Solar neighborhood. 
Stars are removed from the simulation when their distance from the density 
centre exceeds twice that of the tidal radius of the cluster. 
The metallicity of the stars in the two simulations starting with $100\,000$ stars 
(K100-5 and K100-10) was set to be $Z = 0.001$ while both the K24-50 
and K50-20 simulations were assigned solar metallicity ($Z = 0.02$).

\section{Binary Period Distributions}
\label{s:bindst}

\begin{figure}
\plotone{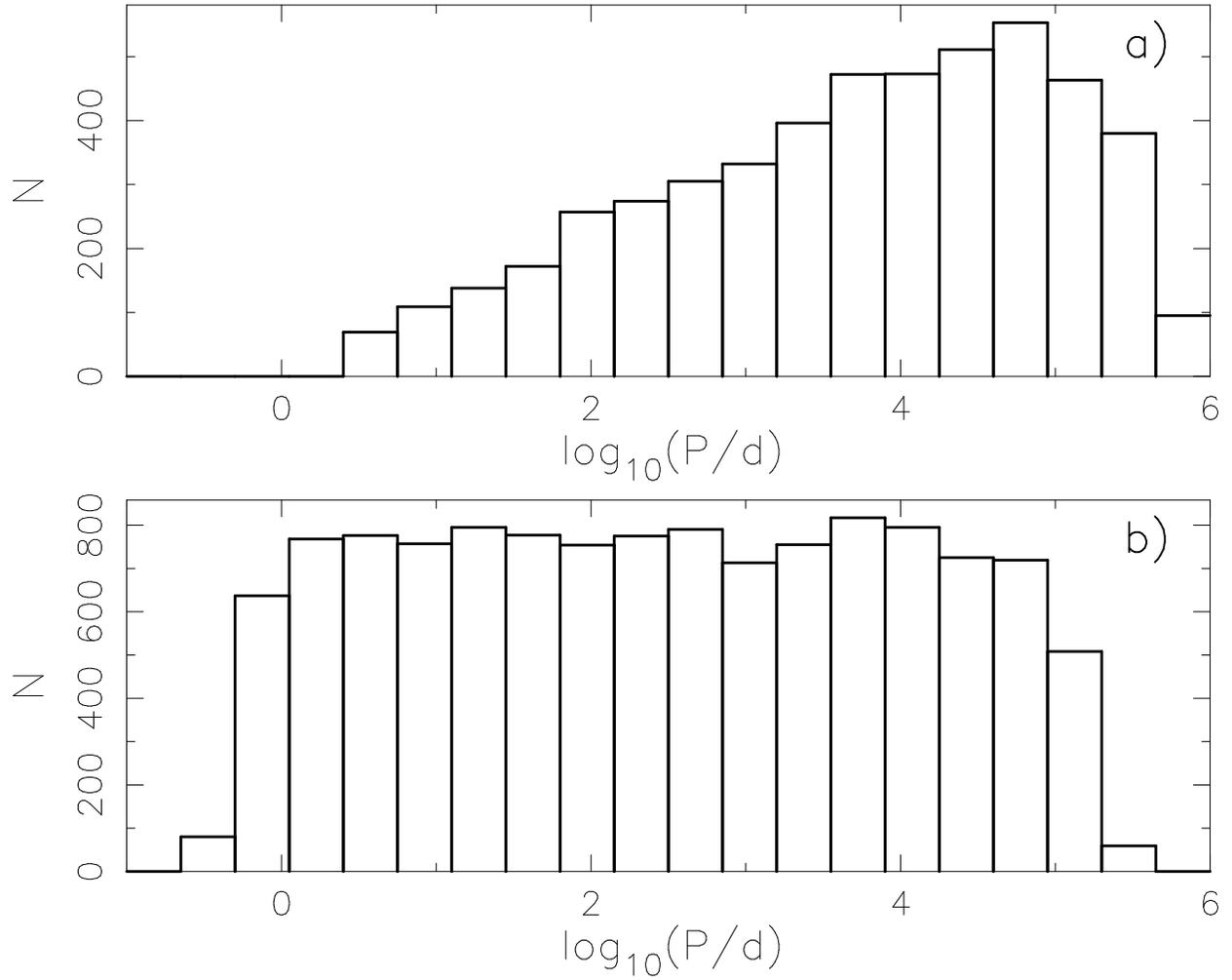}
\caption{
Period distribution of the primordial binary populations in: 
a) the K100-5 simulation (starting with $5\,000$ binaries); and 
b) the K24-50 simulation (starting with $12\,000$ binaries).  
\label{f:fig2}}
\end{figure}

The orbital separations of the $5\,000$ primordial binaries in the K100-5 
simulation (Shara \& Hurley 2006) were drawn from the 
log-normal distribution suggested by Eggleton, Fitchett \& Tout (1989) with a peak at $30\,$au. 
This distribution is based on the properties of doubly-bright visual binaries in 
the Bright Star Catalogue (Hoffleit 1983) 
and is in agreement with the survey data of Duquennoy \& Mayor (1991) 
for binaries in the solar neighbourhood 
-- although the latter observations do not rule out a flat distribution. 
Orbital eccentricities of the primordial binaries were assumed to follow a 
thermal distribution (Heggie 1975). 
In the K100-5 model the initial separation distribution was capped at $100\,$au. 
With a half-mass radius of $6.7\,$pc for the initial model the 
hard/soft binary boundary is at about $30\,$au. 
Thus the maximum of $100\,$au 
excludes only the softest binaries from the distribution.  
Binaries with an initial pericentre distance less than five times the radius 
of the primary star were rejected in the setup of the model 
-- for binaries closer than this it is assumed that interaction during 
the formation process and on the Hayashi track would lead to collision. 
Rather than enact such a collision we simply choose another set of binary parameters 
from the distributions. 
In this way the intended primordial binary fraction is preserved. 
The resulting period distribution of the K100-5 model is shown in Figure~\ref{f:fig2}a. 
We see that the distribution is peaked at $10^5\,$d and 
does not extend beyond $10^6\,$d. 
The K100-10 simulation had the same binary setup as that of the K100-5 
model. 
The K50-20 simulation also used the same Eggleton, Fitchett \& Tout (1989) distribution of 
orbital separations but with a cap at $50\,$au. 

Primordial binaries in the M67 (or K24-50) simulation of Hurley et al. (2005) have orbital 
separations drawn from a flat distribution of $\log a$ (Abt 1983). 
An upper cutoff of $50\,$au was applied so that soft binaries were 
not included in the model 
-- with a half-mass radius of $3.9\,$pc the hard/soft binary limit 
for the starting model was about $40\,$au. 
For this model very close primordial orbits were also rejected. 
The corresponding period distribution for the primordial binaries in the 
K24-50 simulation is shown in Figure~\ref{f:fig2}b. 
We note that a goal of the K24-50 simulation was to reproduce the relatively large 
number of blue stragglers observed in M67. 
For this purpose an Eggleton, Fitchett \& Tout (1989) separation distribution was ruled out as it did 
not lead to enough blue straggler production from Case~A mass transfer in 
close binaries. 
Uncorrelated masses of the component stars in binaries were also ruled out 
for the same reason (see Hurley et al. 2005 for details). 

During this work we will be making comparisons to the Monte Carlo 
models presented by Ivanova et al. (2005). 
In their study binary periods were chosen from a uniform distribution in $\log P$  
between the limits of $0.1$ and $10^7\,$d. 
Thus they assumed a wider distribution of primordial binaries. 
If, for example, the Eggleton, Fitchett \& Tout (1989) 
distribution used in the K100-5 simulation was extended 
to include all periods up to $10^7\,$d, rather than being curtailed at 
$100\,$au, the $5\,000$ binaries that make up the distribution 
shown in Figure~\ref{f:fig2}a would represent about 5/6 of the full population. 
So effectively there would be $1\,000$ soft binaries that have been neglected and 
the true primordial frequency would be 6\%. 
One could then assume that these soft binaries were broken-up 
at the very start of the simulation -- although this may not be true for 
soft binaries residing in the less-dense outer regions of the cluster. 
However, we note that there is no evidence that binary periods in 
star clusters should extend as far as $10^7\,$d (Meylan \& Heggie 1997). 

In terms of hard binaries one could argue, for the sake of semantics,  
that in comparison to a population drawn from a uniform distribution of periods 
extending from $0.1\,$d (without restriction) our initial distributions are 
under-sampling the contribution of hard binaries. 
A key point here is that short-period binaries were not excluded 
from the primordial populations of our simulations  by some ad-hoc process. 
Instead the distribution of orbital periods is dictated by using distributions 
borne from observations in combination with accounting 
for pre-main sequence (MS) evolution 
-- before contracting along the Hayashi track the stellar radius of a pre-MS star can be 
a factor of five or more greater than on the ZAMS (Siess, Dufour \& Forestini 2000) 
and birth periods must allow for this (Kroupa 1995). 
Pre-MS evolution was not considered by Ivanova et al. (2005) although they 
did reject systems where one or both stars would initially fill their Roche-lobes 
at pericentre -- this was also assumed in our models.

\section{Results}
\label{s:result1}

\begin{figure}
\plotone{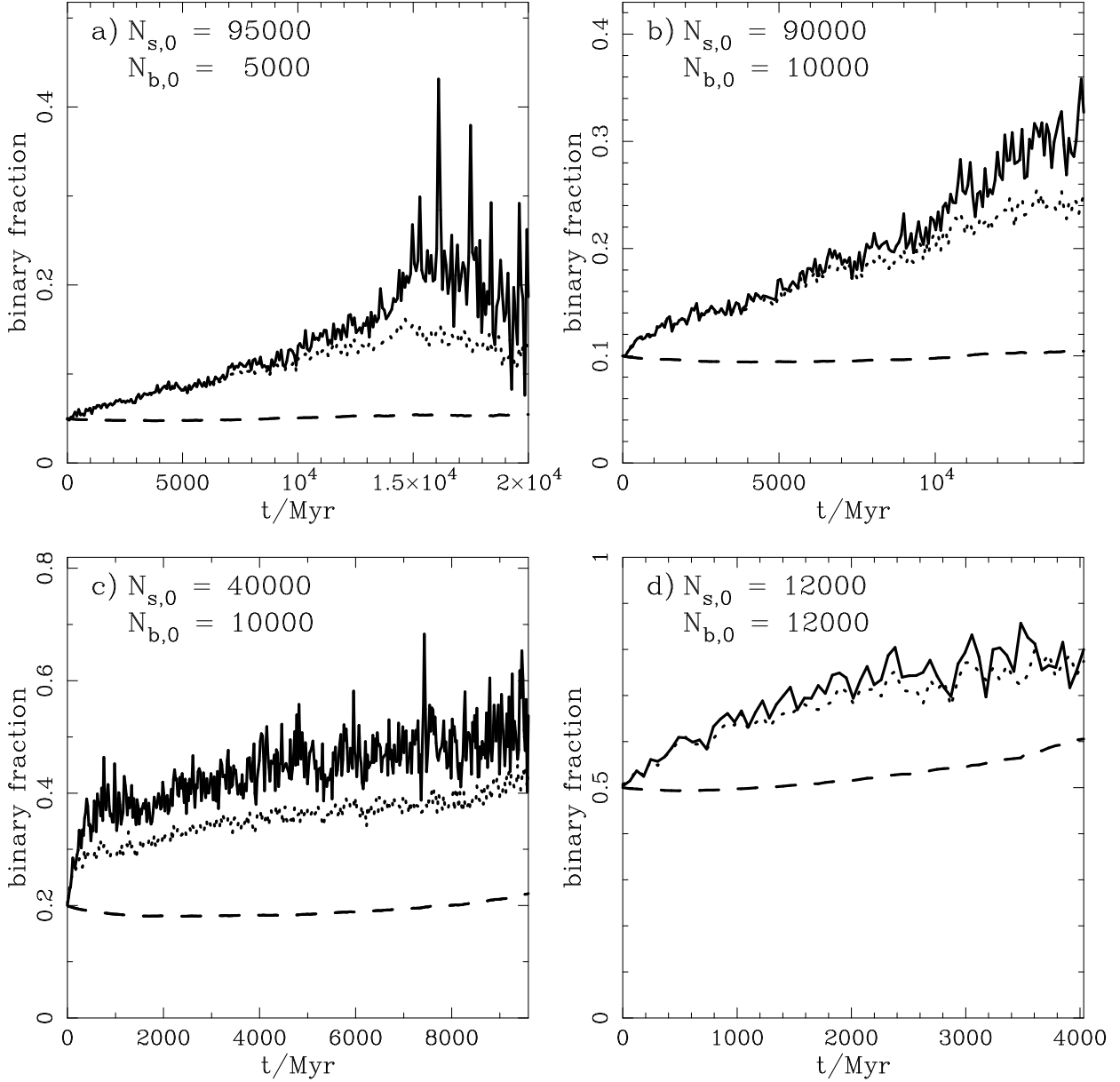}
\caption{
Evolution of the binary fraction in the core (solid line), 
within the 10\% Lagrangian radius (dotted line), 
and for the entire cluster (dashed line). 
Results are shown for the: 
a) K100-5; 
b) K100-10; 
c) K50-20; and 
d) K24-50 simulations 
(see Table~\ref{t:table1} for details). 
\label{f:fig3}}
\end{figure}

In Figure~\ref{f:fig3} we show the evolution of the core binary fraction for the 
four $N$-body simulations introduced above. 
Also shown is the binary fraction within the 10\% Lagrangian radius 
and the overall binary fraction of the model clusters. 

\begin{figure}
\plotone{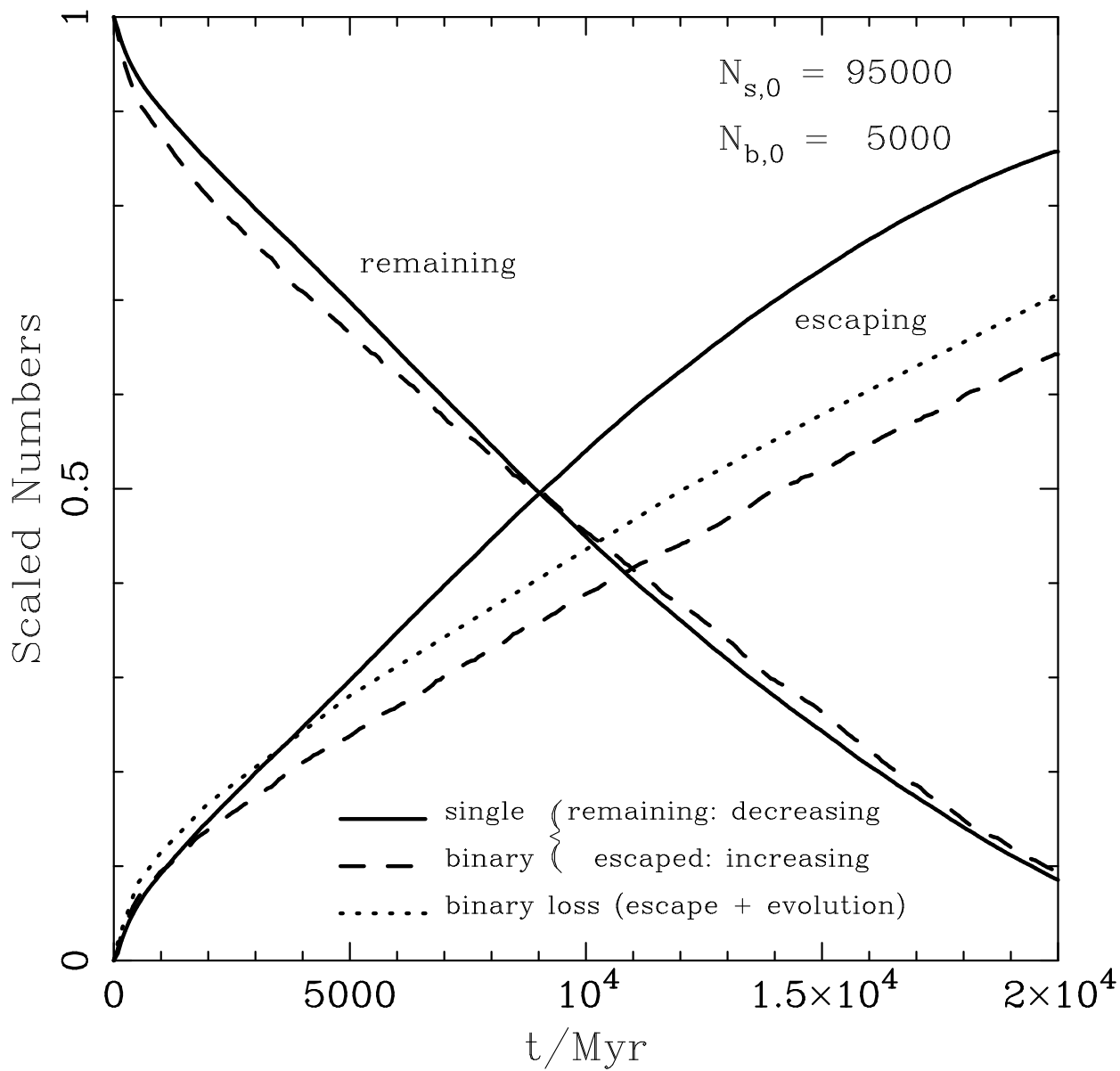}
\caption{
Fraction of single stars (solid line) and binaries (dashed line) 
remaining in the cluster as a function of time 
(lines decreasing from top-left). 
Each population is scaled by the initial number of that 
population. 
Also shown are the fractions of single stars and binaries that 
have escaped from the cluster (lines increasing from bottom-left). 
The dotted line is the combined fraction of binaries lost to escape 
and binary/stellar evolution processes. 
Results are for the K100-5 simulation. 
\label{f:fig4}}
\end{figure}

Except at late times in the K24-50 model, when the cluster has lost more 
than 90\% of its original mass and is nearing dissolution, we see that 
in each case the cluster binary fraction remains close to the primordial value. 
Focussing on the K100-5 simulation, Figure~\ref{f:fig4} shows the fractions  
of single stars and binaries (compared to their respective initial number) 
in the cluster. 
Following on from Figure~\ref{f:fig3}a the fractions are similar at all times 
as expected. 
However, Figure~\ref{f:fig4} also shows the fractions of single stars and 
binaries that have escaped the cluster and we see that from about $2\,$Gyr 
onwards the fractional escape rate of single stars is greater than that of the binaries. 
At the end of the simulation ($20\,$Gyr) the difference is 34\%. 
This is offset somewhat by evolution processes (stellar and binary) that 
destroy binaries (see the dotted line in Figure~\ref{f:fig4}). 
These processes include binaries becoming unbound due to supernova 
mass-loss and/or kicks (only relevant for the first $100\,$Myr of evolution) 
and mass transfer-induced mergers in close binaries. 
The remaining difference is balanced by the destruction of binaries in 
dynamical encounters and this becomes more important as the cluster 
evolves. 
We note that even though the cluster binary fraction is relatively static 
as the cluster evolves the characteristics of the binary population change 
markedly over time with hard binaries favoured at late times. 

Evident from Figure~\ref{f:fig3} is an overall trend for the core binary fraction 
to increase with time, irrespective of simulation type. 
For the core binary population of the K100-5 model we see that this rises 
from an initial 5\% to as high as 40\% around the time that the core-collapse phase is halted.  
After this time the core binary fraction becomes quite noisy owing to the 
small size of the core (see Figure~\ref{f:fig1}) and the small numbers of binaries 
and stars in the core. 
However, the value always remains greater than the initial value. 
We see also from Figure~\ref{f:fig3}a that the binary frequency within the inner 
10\% Lagrangian radius rises to a maximum of 16\% just prior to the end 
of the core-collapse phase. 

It is important to note at this point that we are working with radii derived from 
spherical data whereas observational determinations of binary fractions are 
based on two-dimensional projected data. 
With our models it is possible to test the effect of this discrepancy on our findings. 
If we calculate the 10\% Lagrangian radius for model K100-5 from a two-dimensional 
projection we find that the radius is reduced by about 20-40\% across the evolution 
(the choice of projection axis does not affect this result). 
This is consistent with the expectation suggested by Fleck et al. (2006). 
A similar relationship is reported by Baumgardt, Makino \& Hut (2005) in 
that the half-light radius (calculated from projected data) is approximately half 
the size of the half-mass radius (calculated from spherical data). 
However, the binary fraction within the projected 10\% Lagrangian radius of 
our K100-5 model is almost indistinguishable from that of the result shown 
in Figure~\ref{f:fig3}a (the dotted curve). 

\begin{figure}
\plotone{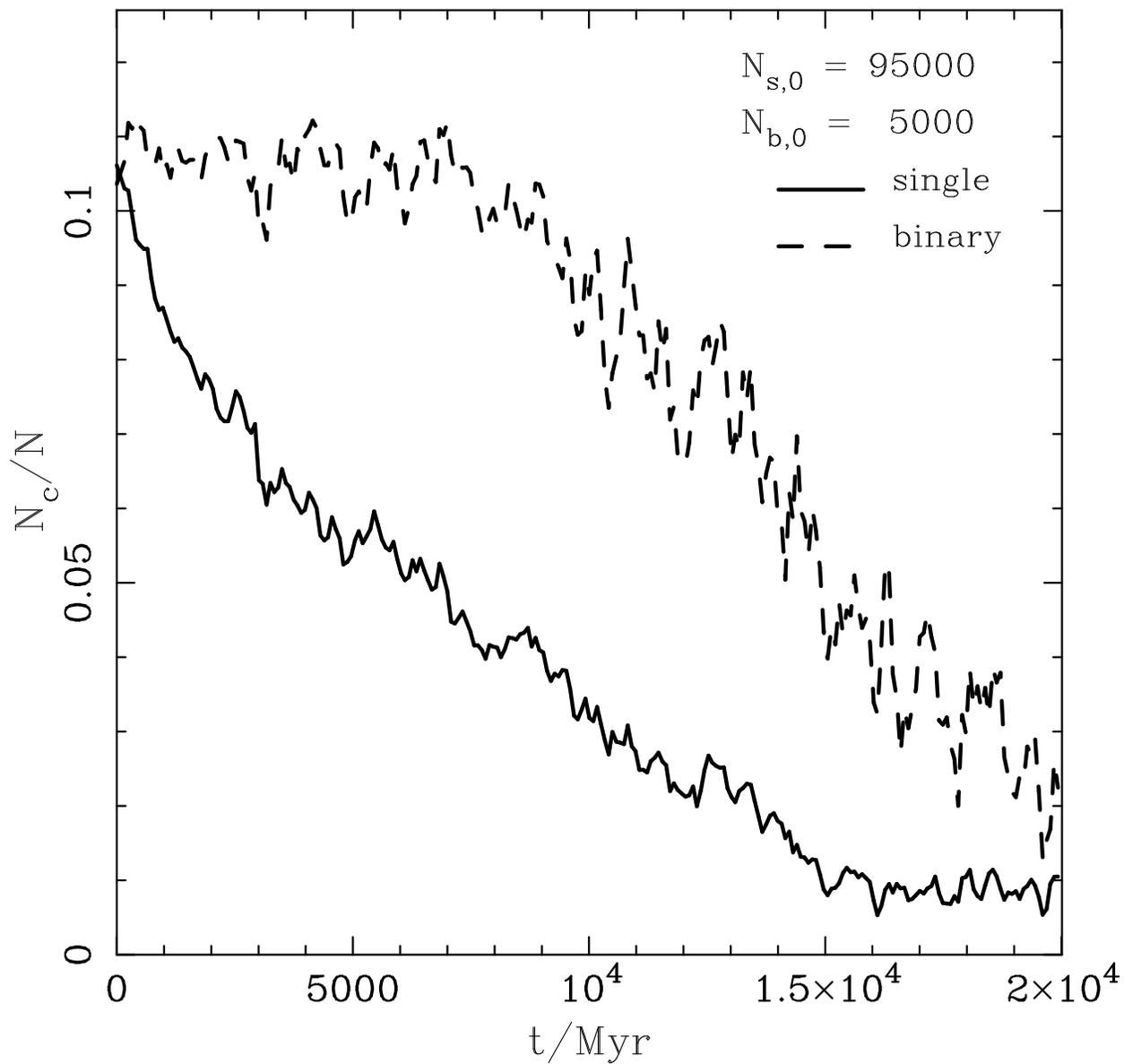}
\caption{
Number of single stars in the core as a fraction of the number 
of single stars in the cluster (solid line) and number of binaries 
in the core as a fraction of the number of binaries in the cluster 
(dashed line). 
Results are for the K100-5 simulation. 
\label{f:fig5}}
\end{figure}

We now aim to understand the processes underlying the evolution of the 
core binary fraction of star clusters, focussing again on the K100-5 simulation. 
Figure~\ref{f:fig5} shows the number of single stars and binaries in the core, 
relative to their total number in the cluster, as the cluster evolves. 
For the first $10\,$Gyr of evolution the ratio of binaries in the core to binaries 
in the cluster is fairly static -- roughly 1 in 10 binaries is in the core. 
However, the ratio of single stars found in the core is decreasing sharply 
over the same timeframe and thus single stars are being lost from the core 
at a greater rate than from the cluster in general 
(comparing Figs.~\ref{f:fig4} and \ref{f:fig5}). 
From $10\,$Gyr onwards the ratio of binaries in the core also decreases. 
This corresponds to a period of increasing core density: 
prior to $10\,$Gyr the core density of stars hovers around the 
$10^2 \, {\rm stars} \, {\rm pc}^{-3}$ mark but from $10-15\,$Gyr 
it increases by an order of magnitude.  
The binary fraction continues to rise in the core over this period indicating that 
single stars continue to be lost from the core at a greater rate than binaries. 
We note that mass loss from stellar evolution is reduced considerably at 
this stage compared to earlier in the cluster lifetime when more massive stars 
were present. 

\begin{figure}
\plotone{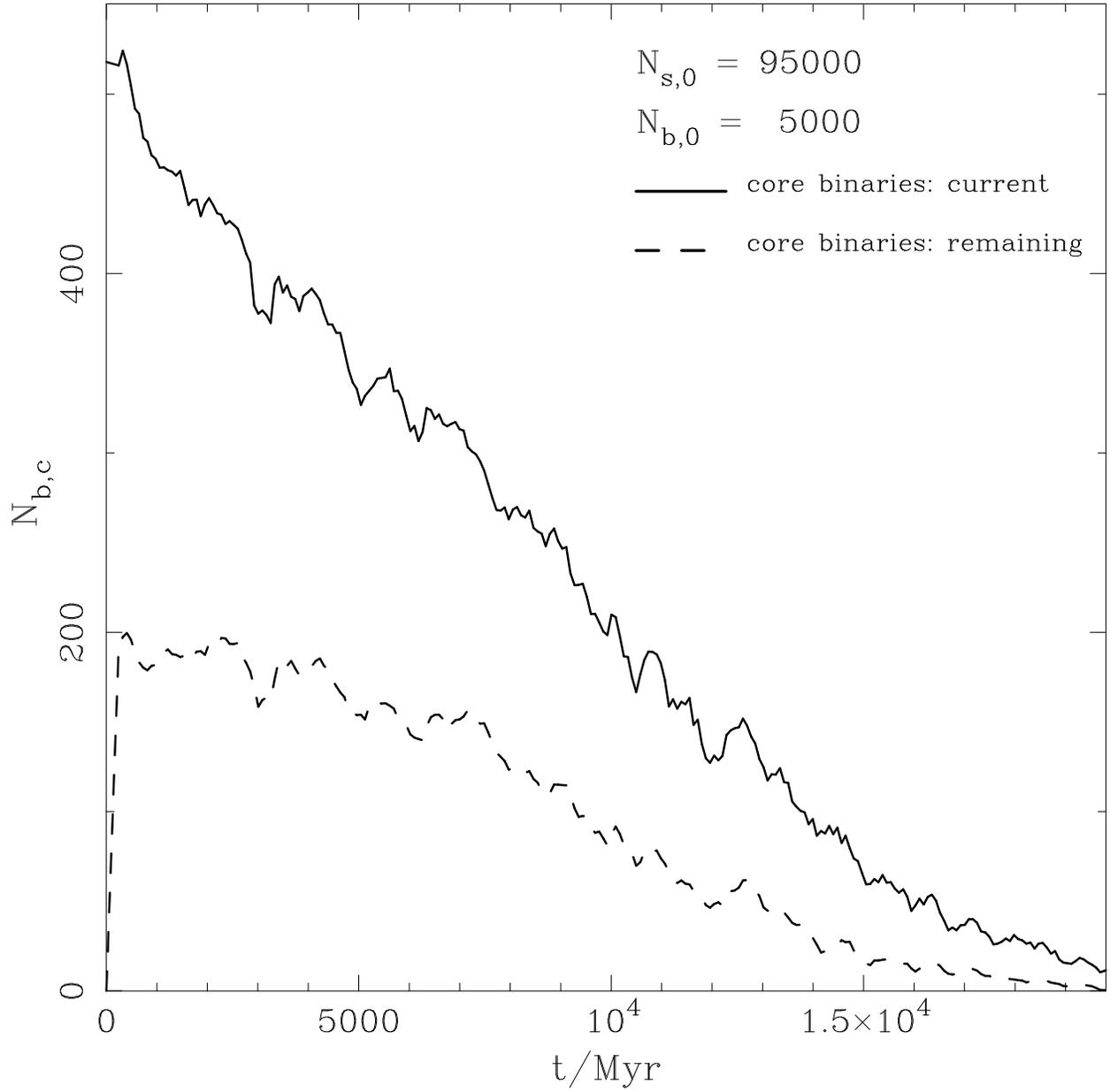}
\caption{
Number of core binaries as a function of time (solid line). 
Also shown at each time is the number of binaries that 
have remained in the core from the previous sampling (dashed line). 
Results are for the K100-5 simulation 
and the data are sampled every $80\,$Myr. 
\label{f:fig6}}
\end{figure}

Figure~\ref{f:fig6} confirms that the {\it number} of binaries in the core is decreasing 
with time even though the {\it binary fraction}, $f_{b, {\rm c}}$, is increasing. 
We also see from this figure that at least half of the binaries in the core at 
any time were not present in the core the last time the population was sampled 
(this is done at intervals of $80\,$Myr). 
So the core binary population is by no means static as many binaries are being 
created/destroyed, or moving in and out of the core, on the $80\,$Myr timescale. 
It is important to note for comparison that the relaxation time in the core is 
approximately $200\,$Myr initially and decreases to about $50\,$Myr 
at late times. 
{\it Individual binaries in cluster cores are both
promiscuous and mobile -- transient residents.}

\begin{figure}
\plotone{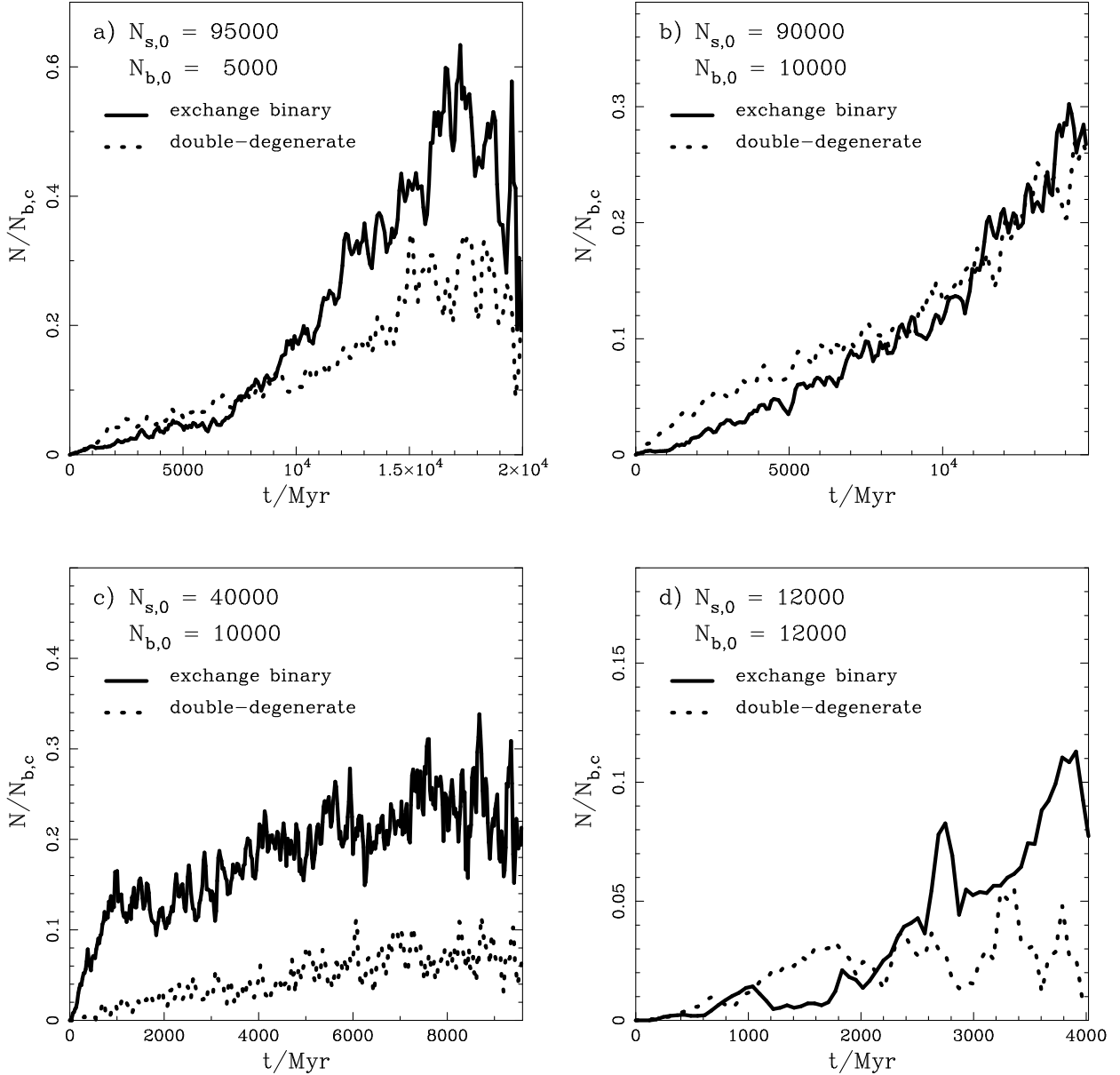}
\caption{
Fraction of binaries in the core that were created in 
an exchange interaction (solid line) and fraction of 
core binaries that contain two degenerate stars 
(dotted line). 
Results are shown for the: 
a) K100-5; 
b) K100-10; 
c) K50-20; and 
d  K24-50 simulations 
(as described in Table~\ref{t:table1}). 
\label{f:fig7}}
\end{figure}

In Figure~\ref{f:fig7} we examine the fraction of core binaries that were created 
in exchange interactions. 
These are short-lived 3- and 4-body gravitational encounters where a star 
is exchanged  into an existing binary displacing one of the members of that 
binary (Heggie 1975). 
Thus it is a process by which primordial binaries can be destroyed and 
replaced by new {\it dynamical}, or {\it exchange}, binaries. 
We see from Figure~\ref{f:fig7}a that these non-primordial binaries come to 
dominate the core population towards the end of the core-collapse phase 
in the K100-5 simulation. 
Figure~\ref{f:fig7}a also shows that the double-degenerate binary content 
increases steadily in the core with time and comprises about 30\% of the 
core binaries subsequent to the completion of the core-collapse phase. 
In Figure~\ref{f:fig7}b we see that the exchange binary content in the 
core of the K100-10 model does not reach the heights of the K100-5 model. 
Presumably this is a consequence of the lower core-density of the K100-10 model. 
The fraction of double-degenerate binaries is similar -- any decrease in 
double-degenerate production via dynamical means in the K100-10 model is 
compensated by the increased number of primordial binaries. 
The fraction of exchange binaries in the core of the K24-50 simulation 
(Figure~\ref{f:fig7}d is comparatively low whereas the K50-20 simulation 
(Figure~\ref{f:fig7}c exhibits a much larger fraction. 
Clearly there is a positive correlation between core density and the fraction of 
exchange binaries in the core.  

\begin{figure}
\plotone{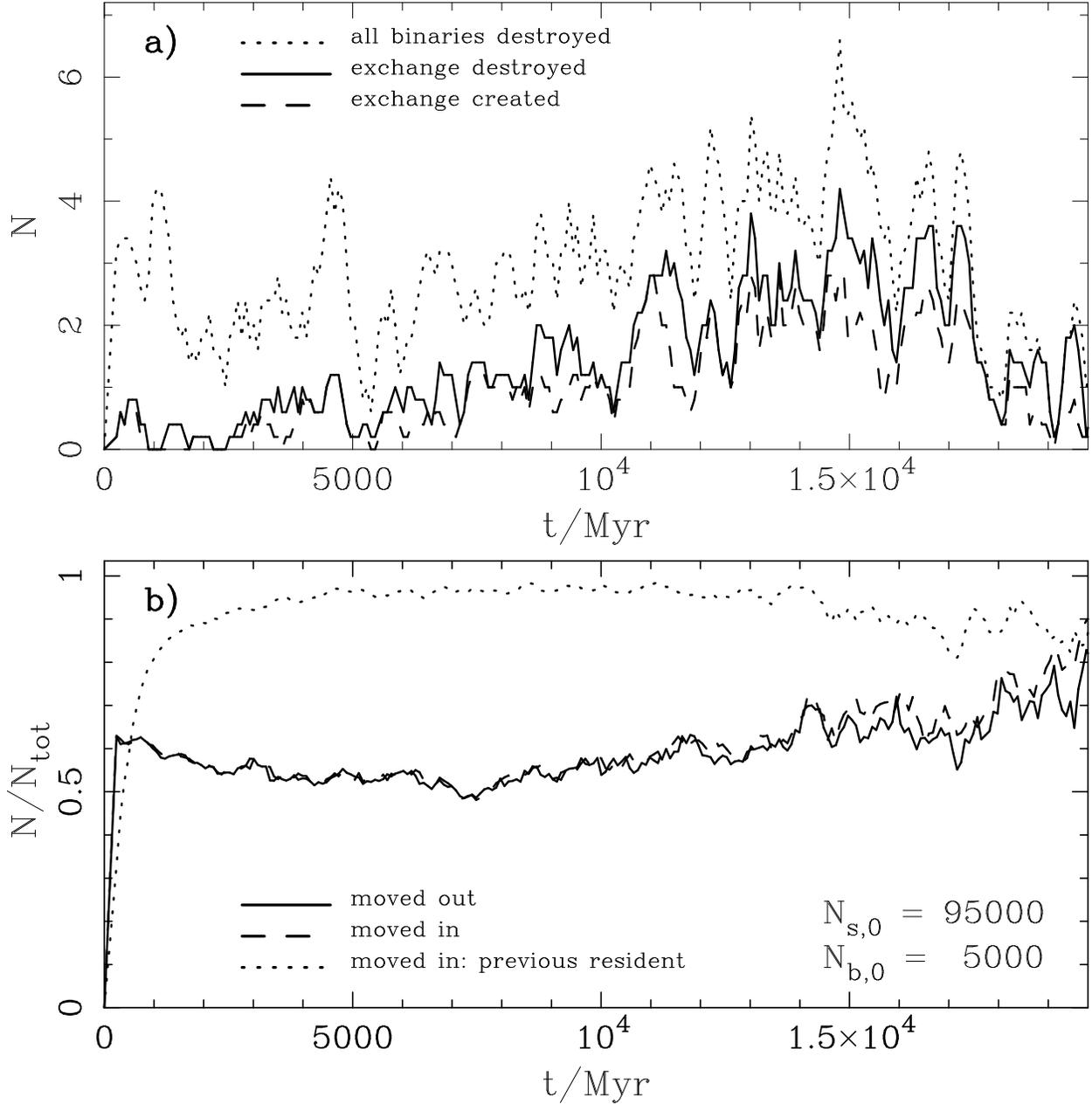}
\caption{
Statistics regarding core binaries across intervals of $80\,$Myr 
as the K100-5 model cluster evolves. 
Shown are: 
a) number of binaries destroyed in an exchange interaction 
occurring in the core (solid line), 
number of binaries created in exchange interactions in the 
core (dashed line) 
and the number of binaries destroyed by any means (dotted line); and 
b) fraction of binaries that have moved out of the core but 
remained in the cluster (solid line: as a fraction of the number 
of binaries in the core at the start of the interval), number of 
binaries that have moved into the core (dashed line: as a fraction 
of the number of binaries in the core at the end of the interval) 
and the fraction of binaries entering the core that have previously 
resided in the core (dotted line). 
Note that the data have been moderately smoothed -- 
over a width of three bins (or  $240\,$Myr). 
Further smoothing would hide the naturally irregular behaviour of the binary 
destruction/creation processes. 
\label{f:fig8}}
\end{figure}

Figure~\ref{f:fig8}a shows the number of binaries created and destroyed 
in exchange interactions occurring in the core in intervals of $80\,$Myr. 
Also shown is the number of core binaries destroyed by all processes 
(exchanges, orbital perturbations, supernovae, mergers) in each interval. 
The key point to note here is that on average exchange interactions are 
creating as many binaries as they are destroying. 
For the entire cluster there were $1\,024$ binaries destroyed in exchange 
interactions during the simulation and $933$ binaries created. 

Figure~\ref{f:fig8}b looks at the movement of binaries in and out of the core 
as the cluster evolves. 
Across each $80\,$Myr interval it shows the fraction of core binaries that 
move out of the core during the interval and the fraction of binaries that have 
moved into the core during the interval. 
We see that the inwards and outwards fluxes are equal. 
Also shown is the fraction of binaries entering the core that have previously 
been in the core -- most binaries that leave the core eventually revisit it. 
We see a pattern where binaries move outwards across the core boundary 
owing to recoil velocities from gravitational encounters, or as a result of 
the shrinking core. 
The core binary population is then replenished by binaries sinking 
inwards owing to mass-segregation effects. 
In the discussion below we will refer to this pattern as {\it binary convection}. 
We note that binaries on radial orbits with a moderate to high eccentricity 
will also make an apparent contribution to this process. 

\begin{figure}
\plotone{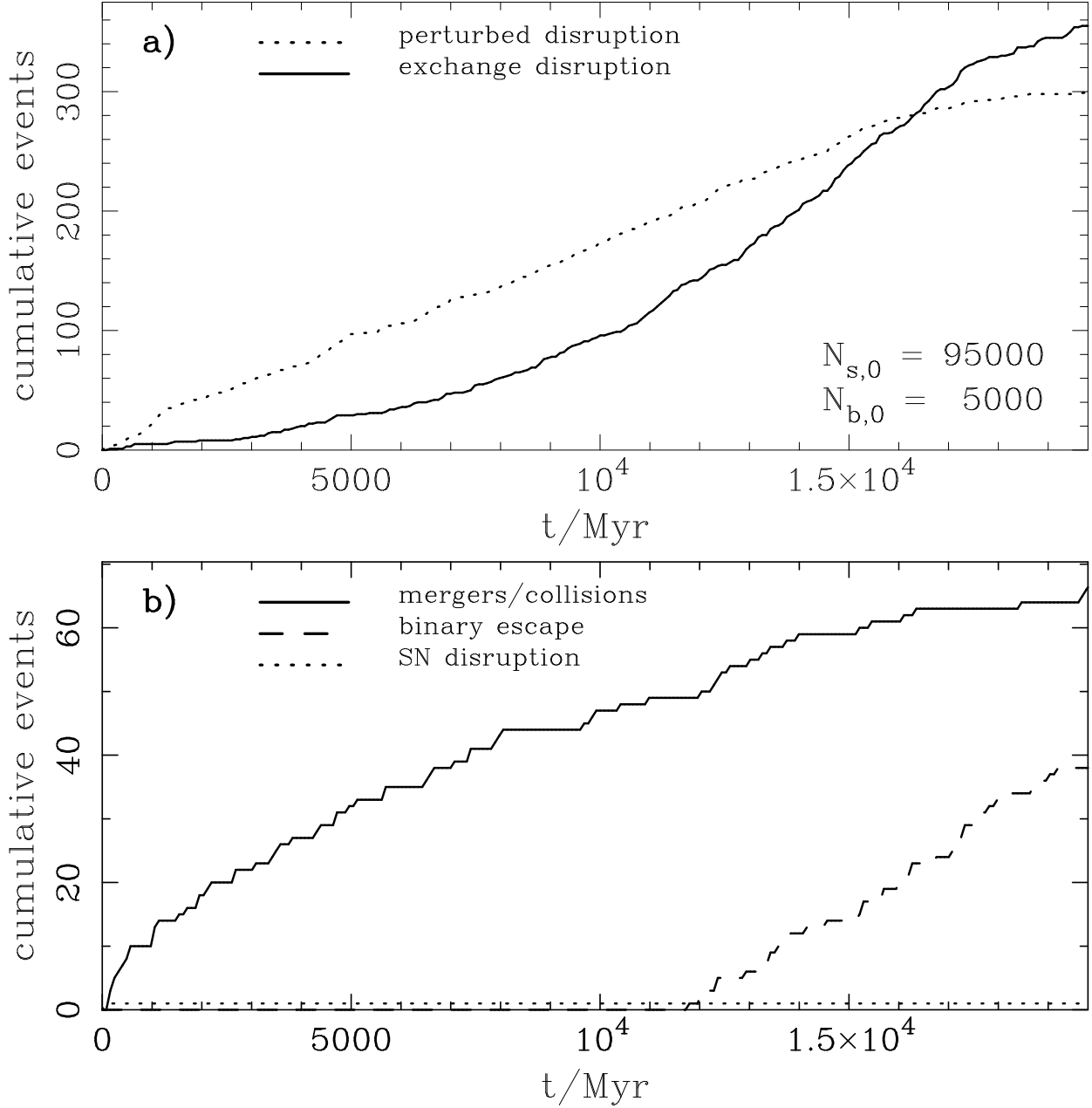}
\caption{
Cumulative numbers of events that lead to the destruction 
of binaries in the core. 
Shown are: 
a) binaries broken-up in exchange encounters (solid line) 
and binaries broken-up owing to orbital perturbations (dotted line); 
b) binaries that were ejected from the core and escaped from the 
cluster (dashed line), binaries broken-up as a result of supernovae  
explosions (dotted line) and binaries in which the stars merged 
(solid line) -- this includes stellar evolution induced mergers and 
collisions at periastron in highly eccentric binaries. 
Results are for the K100-5 simulation. 
\label{f:fig9}}
\end{figure}

An analysis of binary disruption for the K100-5 simulation is given 
in Figure~\ref{f:fig9} in terms of cumulative events. 
Exchange interactions and orbital perturbations from nearby stars are 
by far the dominant causes of binary disruption and these are shown 
in the top panel. 
We see that perturbation events are more likely at early times in the 
evolution but, as soft binaries disappear and the binary population 
becomes skewed towards hard binaries, exchange events eventually 
overtake perturbations as the major cause of disruption. 
However there is an important distinction to make between these two 
types of event. 
Exchange interactions are counted as a disruption event in Figure~\ref{f:fig9}a 
even if the event also leads to the creation of a new binary and as we 
have seen in Figure~\ref{f:fig8}a this is more than likely. 
On the other hand if a binary is broken up owing to an orbital perturbation 
(also known as a fly-by) there is no possibility of a replacement binary 
being created in the event. 

The lower panel of Figure~\ref{f:fig9} shows the number of binaries that 
were ejected from the core and escaped the cluster. 
There is a sharp correlation between the incidence of escape and the 
increase in core density after $10\,$Gyr. 
Even so, the total number of 
binaries lost owing to this process remains an order of magnitude less 
than either perturbation or exchange disruption. 
There is an initial burst of stellar/binary evolution induced mergers in 
short-period primordial binaries followed by a gradual depletion of binaries 
owing to this process and collisions in highly eccentric binaries. 
The cluster had a total of 287 binaries that experienced either a merger or 
an internal collision and 67 of these events occurred in the core. 
We also see from Figure~\ref{f:fig9}b that supernova events do not make 
a meaningful contribution to depletion of the core binary population. 

\begin{figure}
\plotone{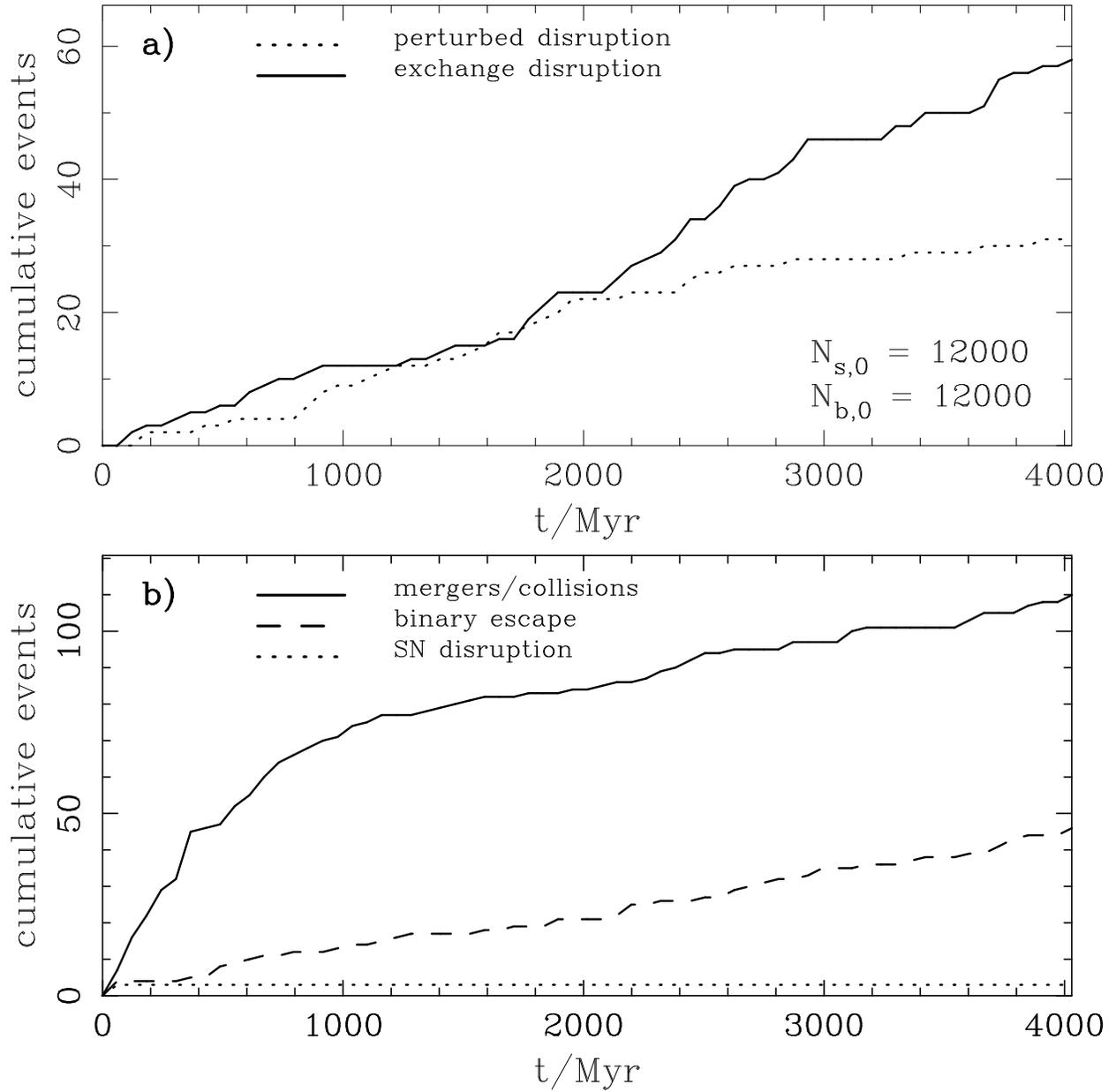}
\caption{
As for Figure~\ref{f:fig9} but for the K24-50 simulation. 
\label{f:fig10}}
\end{figure}

Figure~\ref{f:fig10} repeats  Figure~\ref{f:fig9} for the K24-50 simulation. 
In this simulation mergers and collisions are the most likely cause 
of core binary loss. 
This is linked to the increased primordial binary fraction and decreased 
core density, compared to the K100-5 simulation. 
For similar reasons exchange disruption is more likely than perturbed 
disruption over the course of the evolution. 
In fact in this simulation even the loss of binaries from the core as a result 
of escape is greater than that from perturbed break-up. 
A key distinction between the K24-50 and K100-5 simulations is that in the 
K24-50 case the ratio of binary destruction to creation in exchanges is 
3:1 whereas it was close to 1:1 for the K100-5 simulation. 

The effect of a substantial primordial binary population on the evolution 
of open clusters has been documented in the past 
(McMillan, Hut \& Makino 1990, for example and see Meylan \& Heggie 1997 for a review). 
The main results are that in comparison to simulations without primordial 
binaries the core-collapse phase of evolution is less dramatic and the cluster 
lifetime is reduced. 
Little has been done on this subject for globular clusters to date primarily 
because direct simulations have not been possible. 
However, our simulations starting with $100\,000$ stars can start to 
shed some light on the expected behaviour. 
We see from Table~\ref{t:table1} that increasing the primordial binary 
frequency from 5\% (K100-5 simulation) to 10\% (K100-10) does not 
reduce the cluster half-life significantly. 
In contrast the K24-50 simulation with 50\% binaries has a half-life of $2\,060\,$Myr 
while a comparable simulation of $30\,000$ single stars with no primordial binaries has 
a half-life of $3\,600\,$Myr. 
As noted in Hurley \& Shara (2003) the presence of a large number of primordial 
binaries in an open cluster leads to an enhanced rate of escaping stars via recoil 
velocity kicks obtained in 3-body interactions. 
In comparison, the K100-5 and K100-10 clusters have deeper potential wells 
and also the change in binary fraction between the two models is much less 
than for the open cluster example. 
So a sharp change in the escape rate is not to be expected. 

\begin{figure}
\plotone{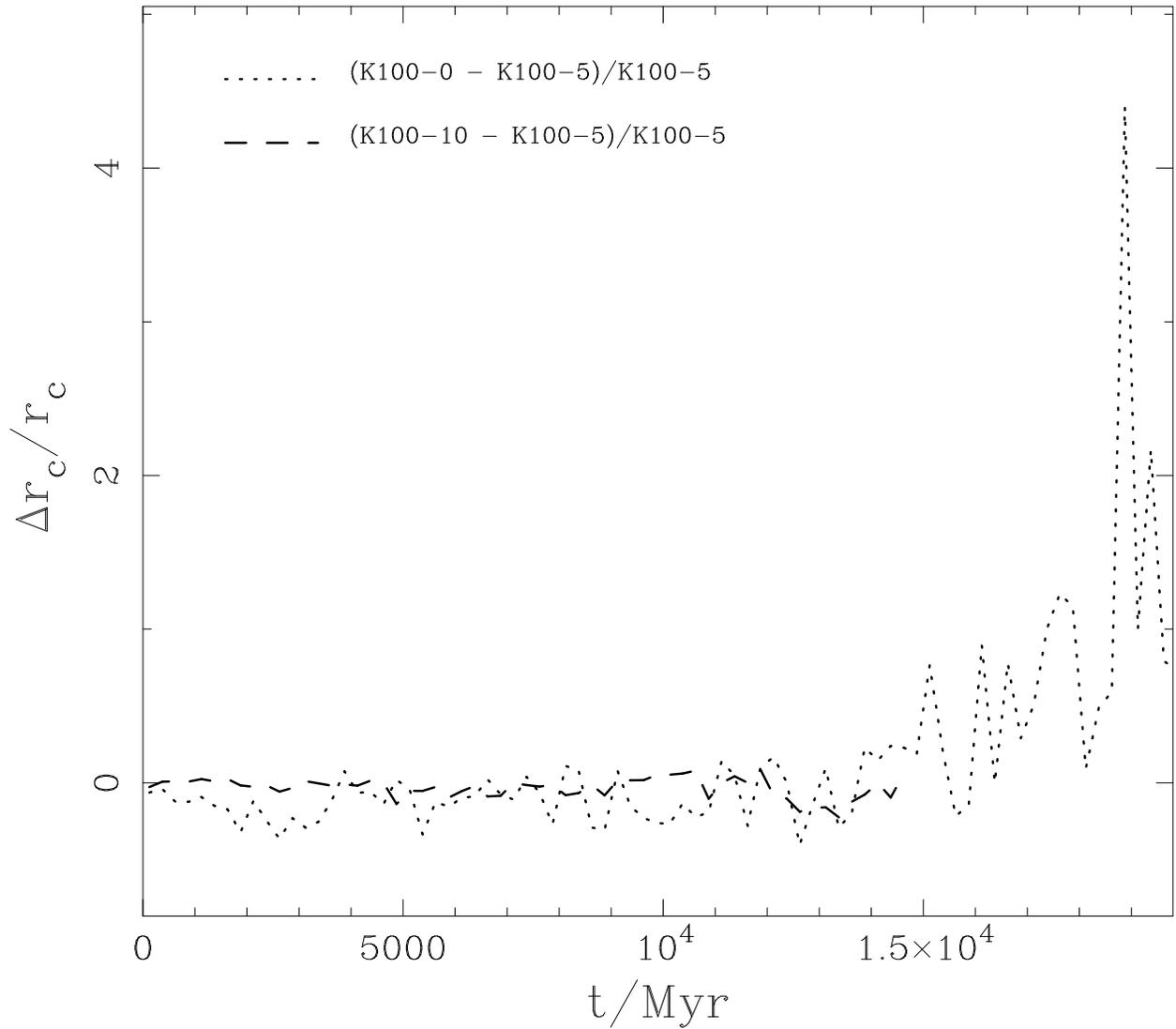}
\caption{
Comparison of core-radius evolution for models starting with 
$100\,000$ stars. 
The K100-5 simulation is taken as a reference model and shown are 
differences between the core radius of this model and models 
starting with 0\% (dotted line) and 
10\% primordial binaries (K100-10: dashed line).  
The difference is scaled by the core radius of the K100-5 model. 
Note that for each simulation the core radius used is the average core radius in 
a $250\,$Myr interval.  
\label{f:fig11}}
\end{figure}

Figure~\ref{f:fig11}  shows that the core radius evolution of the K100-5 and 
K100-10 simulations is similar up to $15\,$Gyr 
(when the K100-10 simulation was stopped).  
We note however that the core density of the K100-10 model at this time is only 
half that of the K100-5 model. 
So the presence of additional primordial binaries has reduced the 
number density of stars in the core. 
Also in Figure~\ref{f:fig11} we compare the core radius evolution of a $100\,000$ star 
simulation with no primordial binaries (a K100-0 model). 
Here we see that the core radius evolution is slightly more irregular  
but overall the evolution is once again similar up to $15\,$Gyr. 
After core-collapse has been halted the situation is different as the single star model experiences 
a fluctuating, and generally increasing, core-radius while the core-radius of the 
K100-5 model remains approximately constant (see Figure~\ref{f:fig1}a). 
The K100-0 model has a greater core density than the K100-5 model at the end of 
the main core-collapse phase. 
The fluctuating core-radius of the K100-0 model in the post-core-collapse phase is 
indicative of the core bounce and subsequent oscillations expected for 
such a model -- these phenomena are more pronounced for models without 
primordial binaries (see the related discussion in Heggie \& Hut 2003 
and Heggie, Trenti \& Hut 2006). 

\begin{figure}
\plotone{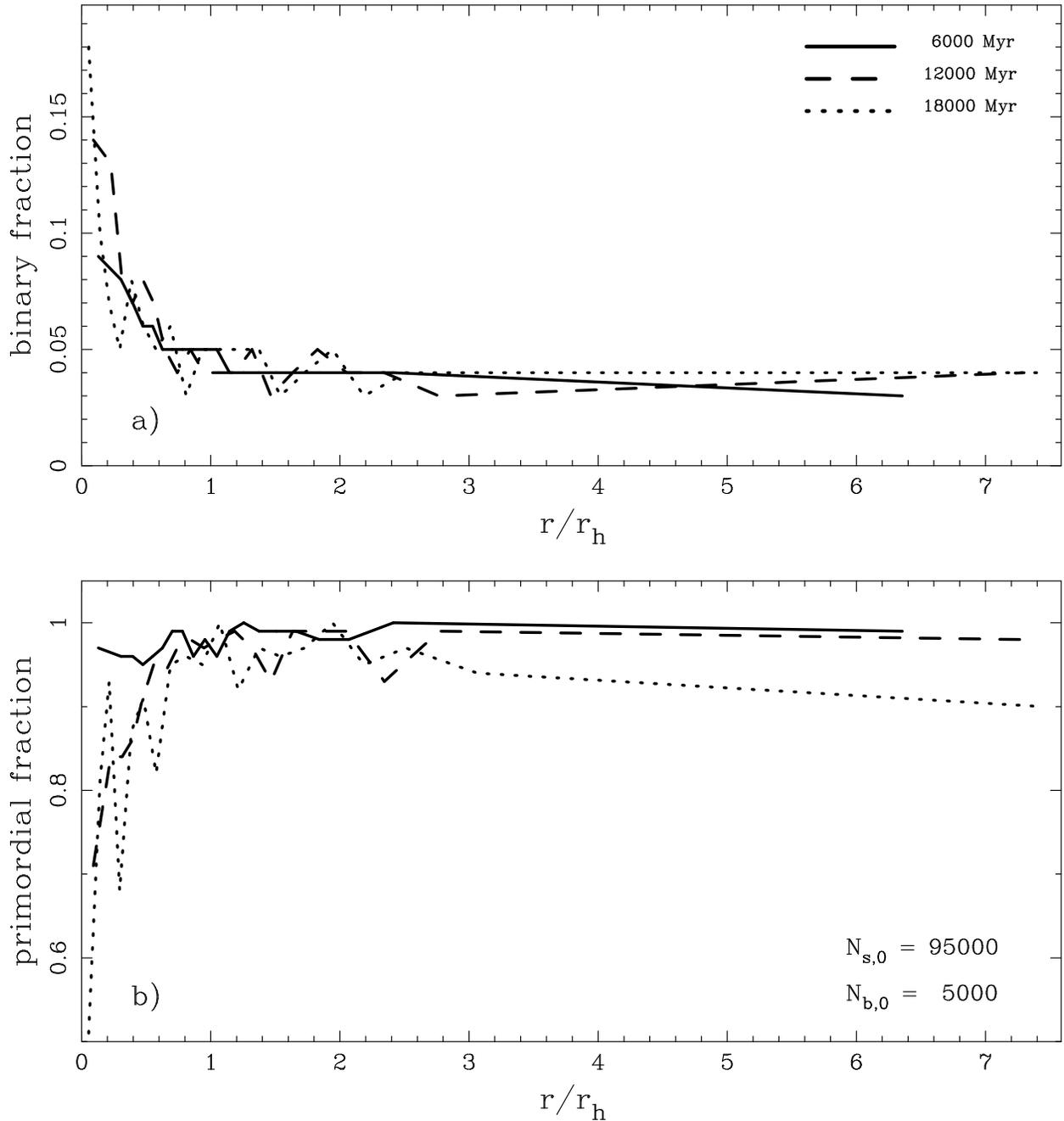}
\caption{Binary data as a function of radial position for the K100-5 model. 
Shown at times of $6\,$Gyr (solid line), 
$12\,$Gyr (dashed line) and $18\,$Gyr (dotted line) are: 
a) distribution of binary fraction; 
b) fraction of binaries that are primordial. 
At each time there are twenty radial bins each containing the same mass, 
i.e. corresponding to Lagrangian radii incremented by 5\%. 
Thus the core is not resolved. 
\label{f:fig12}}
\end{figure}

In Figure~\ref{f:fig12} we investigate the radial distribution of the K100-5 binary 
population at times of $6$, $12$ and $18\,$Gyr, i.e. before, during and 
after the deep core-collapse phase. 
We see that outside of the half-mass radius the binary fraction is 
effectively constant with radius and changes little with time. 
The binary population in this region is also dominated by primordial 
binaries -- exchange binaries are unlikely to be found outside of the 
half-mass radius. 
Within the half-mass radius the binary fraction rises sharply towards 
the centre of the cluster and binaries become more centrally concentrated 
as the cluster evolves. 
Note that the inner radial bin corresponds to the inner 5\% Lagrangian 
radius so the core is not resolved in Figure~\ref{f:fig12}.

\section{Discussion}
\label{s:disc}

Our $N$-body results clearly show that the core binary fraction of an evolved star cluster 
is expected to be greater than the primordial binary fraction. 
We see this behaviour in each of the models presented and at all times 
in the evolution. 
The most striking case is our main model (K100-5) 
which started with $95\,000$ single stars and $5\,000$ binaries and
experienced a factor of eight increase in the core binary fraction after $16\,$Gyr of 
evolution (coinciding with the end of the main core-collapse phase). 

At face value our results appear to be in 
clear contradiction to the Monte-Carlo results recently presented by 
Ivanova et al. (2005). 
Their main reference model has a primordial binary fraction $f_{\rm b,0} = 1.0$ and 
a stellar density of $n_{\rm c} = 10^5 \, {\rm stars} \, {\rm pc}^{-3}$. 
Processes such as exchange interactions, orbital perturbations, binary evolution 
and mass-segregation are included and the model is reduced to $f_{\rm b,c} = 0.095$ 
at $14\,$Gyr. 
Ivanova et al. (2005) then repeat the simulation with $f_{\rm b,0} = 0.5$ and end up with 
$f_{\rm b,c} = 0.07$ so, as they note, the relationship between primordial binary 
fraction and final core binary fraction is not linear. 
Coming at this from the other direction our models show a possible 
saturation effect as the primordial binary fraction increases. 
Looking back at Figures~\ref{f:fig3}a and ~\ref{f:fig3}b we see that the core 
binary fraction of the K100-5 model at the $14\,$Gyr mark is ~0.2 (up from $f_{\rm b,0} = 0.05$) 
while it is ~0.3 (up from $f_{\rm b,0} = 0.1$) for the K100-10 model. 
So the simulation with the lower primordial binary fraction has experienced 
the greater relative increase in core binary content. 
Our K24-50 model, which started with $f_{\rm b,0} = 0.5$, has $f_{\rm b,c} =  0.8$ at a similar 
dynamical age so the relative increase is less again. 
These results raise the possibility that decreasing $f_{\rm b,0}$ below 0.5 in the Monte Carlo 
models may lead to conditions where $f_{\rm b,c}$ can increase. 
It is interesting to note that the idealized models presented recently by 
Heggie, Trenti \& Hut (2006) showed saturation effects in the core for initial binary 
frequencies greater than 10\% and also recorded an increase in the core 
binary fraction with time. 

Ivanova et al. (2005) also performed a model to compare with 47 Tucanae. 
This was similar to their main reference model although slightly more dense and 
with an increased velocity dispersion. 
The result for $f_{\rm b,0} = 1.0$ was $f_{\rm b,c} = 0.07$ 
-- this lead to the conclusion that the primordial binary frequencies of globular clusters 
such as 47 Tucanae must have been close to 100\% to explain current observations. 
However, Ivanova et al. (2005) also ran the same simulation with 
$f_{\rm b,0} = 0.75$, 0.5 and 0.25 and reported little or no variation in the final 
core binary fraction. 
It would seem safe to assume that repeating the simulation with 
$f_{\rm b,0} = 0.1$ may give the same result or even an increase in binary fraction. 
This would act to remove any obvious discrepancies between the $N$-body and 
Monte Carlo results. 
We would certainly be interested in seeing the results of a Monte Carlo simulation 
conducted with $f_{\rm b,0} = 0.1$ and a similar setup of the primordial binary population 
as used in this work 
-- much easier than repeating a large $N$-body simulation with 100\% binaries. 

A major distinction between our $N$-body models and the Monte Carlo simulations 
mentioned above is that the stellar density is at least an order of magnitude greater 
in the latter. 
Fortunately, Ivanova et al. (2005) performed a simulation with 
$n_{\rm c} = 10^3 \, {\rm stars} \, {\rm pc}^{-3}$ which facilitates a more direct 
comparison with our K50-20 model which had a similar core density 
throughout the evolution. 
The K50-20 model experienced an almost factor of two increase in core binary fraction as it 
evolved from $0 - 8\,$Gyr. 
The comparable Monte Carlo model showed a reduction in core binary fraction of more than 
a factor of two over the same period. 
So there is an obvious deviation in behaviour. 
Of course there is a large difference in the primordial binary fractions 
(0.2 compared to 1.0). 
The effect of this will be discussed further below. 
However, we note at this stage that the initial hard binary fraction in 
the Monte Carlo model was $\sim 30\%$ (Ivanova, private communication) 
and this rose to 37\% 
-- so the hard binary fraction increases and subsequently the models do show agreement at 
some level. 
Another consideration is the velocity dispersion which is generally around 
$3-4 \, {\rm km} \, {\rm s}^{-1}$ for our models and was set to $10  \, {\rm km} \, {\rm s}^{-1}$ 
for most of the Monte Carlo models. 
However,  Ivanova et al. (2005) did perform two models (D4 and M12) similar in all respects 
except that $\sigma = 10  \, {\rm km} \, {\rm s}^{-1}$ in one and $4.5 \, {\rm km} \, {\rm s}^{-1}$ 
in the other. 
There was no significant difference in the final core binary fractions of these models. 

In Section~\ref{s:bindst} we discussed that in the setup of our models 
we might be neglecting a fraction of soft binaries from the true 
primordial population.  
This results from imposing a maximum initial orbital separation and at most would 
cause the binary fraction to be underestimated by a few per cent. 
Thus we are confident that our choice of initial parameters for the binary populations 
in our models is not affecting the result that the core binary fraction increases as 
a cluster evolves. 
We also note that differences in the setup of primordial binaries between our 
simulations and those of Ivanova et al. (2005) make it difficult to directly compare 
quoted binary frequencies. 
For example, by not accounting for pre-MS stellar radii as we do, Ivanova et al. (2005) 
have a greater relative number of close binaries in their primordial populations. 
Such an excess would result in a greater number of evolution-induced binary mergers. 
If we were to adopt the period distribution and methods used by Ivanova et al. (2005)  
we would need to choose $\sim 11\,000$ binaries in order to recover the $5\,000$ in 
our K100-5 model at birth. 
This gives an effective primordial binary frequency of 11\%, for the sake of comparison. 
The effective primordial binary frequency for the K24-50 simulation would be 80\%. 
Adopting these values, in the worst case scenario, would still {\it not} lead us to conclude that 
the core binary fraction of an evolved cluster is decreased from the primordial value. 

The comparable rates of binary disruption and creation owing to exchanges  
in our K100-5 simulation indicates that 3-body interactions dominated over 
4-body interactions. 
This is because the most likely outcome of a binary-binary encounter is a 
binary and two single stars. 
So a binary is lost from the overall count. 
This is not the case for binary-single encounters where the most likely outcome 
is a binary and a single star, although the pairing of stars in the binary and/or 
the orbital parameters may have changed. 
By contrast, exchange interactions in the K24-50 simulation produced a 
binary disruption rate much higher than the binary creation rate. 
Here we had a much higher proportion of 
primordial binaries and thus binary-binary encounters were more likely. 
Thus, in terms of exchange interactions, increasing the primordial binary fraction 
can lead to a greater rate of binary destruction. 
This would certainly be expected to be true of models with comparable 
stellar densities. 
However, a competing effect comes from the fact that the central density 
is less for simulations with higher primordial binary fractions. 
We certainly see this when comparing our K100-5 and K100-10 models.  
The setup of these models was identical in all respects except for the 
change in primordial binary frequency from 5\% to 10\%. 
The models have similar half-lives and we showed that the core radius evolution 
is also similar. 
So at any particular time in the evolution they are at a comparable dynamical age. 
But there is one clear difference -- the model with twice as many primordial binaries 
has a central stellar density that is a factor of two less. 
This translates to a lower incidence of close stellar encounters and as we saw from 
Figure~\ref{f:fig7} a greatly reduced fraction of exchange binaries in the core. 
Previous simulations, albeit with small-$N$, have indicated that the effects of 
primordial binaries saturate at some level (Wilkinson et al. 2003) so this is not 
necessarily a trend that we expect will continue as the primordial 
binary fraction is increased towards unity. 
However it is certainly significant for clusters with frequencies of 10\% or less. 

Another point to note is that in a 3-body exchange, not only is a binary not lost,  
but also a more massive single star is swapped for a less massive one, 
increasing the likelihood that the single star will be lost from the core via mass-segregation. 
So the exchange process has indirectly increased the core binary fraction. 
The process of binary convection that became evident from Figure~\ref{f:fig8}b also 
is related to mass-segregation and acts to keep the core binary fraction healthy. 
Both single stars and binaries in the core are subject to velocity kicks from 
gravitational encounters. 
These kicks can remove an object from the core and even from the cluster entirely. 
For binaries this is less likely to occur primarily because they are on average 
more massive than single stars. 
Also, the average stellar mass decreases radially outwards in an evolved cluster. 
So if a core binary suddenly finds itself outside of the core it can be expected to be one 
of the more massive objects in its new local environment and thus to quickly 
sink back towards the core. 
We note that we found the movement of binaries inwards and outwards across 
the core boundary, as exhibited by Figures~\ref{f:fig6} and \ref{f:fig8}, to be quite striking. 

Our K100-5 $N$-body simulation creates a realistic model of a moderate-size globular cluster. 
It combines stellar and binary evolution with a self-consistent treatment of the 
cluster dynamics. 
It includes primordial binaries and accounts for the tidal field of the Galaxy. 
Thus it provides us with a solid picture of how such a cluster evolves. 
Single stars escape from the cluster at a greater rate than binaries do 
-- single stars are less massive on average so they are more likely to be tidally 
stripped after segregating to the outer regions of the cluster and also more likely 
to be ejected from the cluster in gravitational encounters. 
However, binaries are also lost from the cluster population owing to supernova 
disruption, evolution-induced mergers and dynamical encounters. 
These effects balance and the ratio of single stars to binaries is similar at all 
times in the evolution. 
As the cluster evolves binaries sink towards the centre and the binary fraction 
increases in the central regions. 
The core radius decreases as core-collapse proceeds and dynamical encounters 
become more prevalent. 
These encounters not only break-up binaries but also create new binaries. 
The cluster evolves to a state where primordial binaries dominate the binary 
population in the outer regions and non-primordial binaries dominate towards 
the centre. 

In the centre of the cluster soft binaries are broken-up as a result of orbital perturbations 
from gravitational encounters. 
Binaries become involved in exchange interactions, primarily three-body, but these 
tend to create as many binaries as they destroy. 
Hard binaries are lost when the components merge as a result of close binary 
evolution or a collision at periastron. 
These are ongoing processes as the cluster evolves. 
At an age of $10\,$Gyr the rate of exchange interactions is greater than that of 
perturbed break-ups and mergers. 
However, perturbed break-ups are the dominant cause of binary loss. 
This is compared to the Monte Carlo model of Ivanova et al. (2005) which 
found that evolutionary mergers were the dominant event at the same age. 
We also find that after $10\,$Gyr, as the core density increases, that binaries 
can be kicked out of the cluster directly from the core. 
Partly as a result of the combination of these processes the number of binaries 
in the core decreases as the cluster evolves. 
Also to blame is the movement of binaries outwards across the core boundary 
owing to the decreasing size of the core and recoil velocities invoked in 
gravitational encounters. 
However, the movement of single stars outwards across the core boundary is greater 
and the net effect is an increase in the core binary fraction. 
This is also helped by binary convection where binaries that were previously 
resident in the core are cycled back in. 

Noting that the typical membership of Galactic globular clusters exceeds $300\,000$ 
stars (Gnedin \& Ostriker 1997, for example) we must ask the question 
-- to what extent can we expect this behaviour to extend to globular clusters 
in general? 
We can start with the ejection rate, $t_{\rm ej}$, of stars from an isolated cluster calculated by 
H\'{e}non (1969) which gives 
$t_{\rm ej} \propto \ln \left( 0.4 N \right) \, t_{\rm rh}$ 
(Binney \& Tremaine 1987). 
Here $t_{\rm rh}$ is the half-mass relaxation timescale and we can relate this 
to behaviour near the core of a cluster if we assume that core-mass scales 
with total mass and that radii do not vary appreciably with cluster mass. 
This indicates that the relative rate of outward binary ejection and inward 
mass-segregation (which occurs on a relaxation timescale) is only weakly 
dependent on the cluster mass. 
If we look in detail at the local relaxation timescale this scales as 
\[ t_{\rm r} \propto \frac{\sigma^3}{\rho \ln \left( 0.4 N \right)} \] 
(Davies, Piotto \& De Angeli 2004, as derived from Binney \& Tremaine 1987) 
where $\sigma$ is the velocity dispersion of the cluster stars and $\rho$ is 
the mass density. 
We can take $\sigma \propto \sqrt{M/r_{\rm h}} \propto M_{\rm c}^{1/2}$ and 
$\rho \propto M_{\rm c}$, using the above assumptions, to show that 
$t_{\rm r} \propto M_{\rm c}^{1/2} / \ln \left( 0.4 N \right)$. 
Here $M_{\rm c}$ is the cluster core-mass, $M$ is the total cluster mass and $r_{\rm h}$ 
is the half-mass radius. 
The timescale for a typical binary in the core of a globular cluster to have a close 
encounter with another star scales as 
\[ t_{\rm enc} \propto \frac{\sigma}{n} \] 
(Davies, Piotto \& De Angeli 2004) where $n$ is the number density and $n \sim \rho$ 
if the average stellar mass is of the order of $M_\odot$, as it is in an evolved cluster core. 
This gives us $t_{\rm enc} \propto M_{\rm c}^{-1/2}$. 
To escape the core a binary must acquire a boost in energy of the order of 
$G \, M_{\rm c} / 2 \, r_{\rm c}$ (where $G$ is the Gravitational constant). 
So, assuming that the average energy imparted in an encounter does not vary strongly 
with mass, we have $t_{\rm ej} \propto M_{\rm c}^{1/2}$. 
This rather simplified analysis returns H\'{e}non's result and shows that as $M$ (or $N$) 
increases there will be relatively less binary convection as both the ejection and relaxation 
timescales increase. 
However, the effect on the observed core binary fraction can be expected to be minimal. 

We cannot definitively use our results to make predictions regarding globular clusters such as 
47 Tucanae because the central density in these clusters is at least an 
order of magnitude higher than that reached by our models.  
However, we note that our model with the highest core density 
showed the greatest increase in core binary fraction. 
Furthermore, we have considered a range of cluster types.   
{\it It does not appear, from our simulations, that an initial binary fraction anywhere 
near as high as 100\% is required to give a core population of 20\% or less at later times.} 
We also note that proper-motion cleaned colour-magnitude diagrams recently presented 
for NGC$\,6397$ (Richer et al. 2006) and M4 (Richer et al. 2004)  
show a distinct lack of binaries in regions outside of the cluster centre 
-- this cannot be reconciled with a large primordial binary population.

\section{Summary}
\label{s:summ}

We have presented a range of simulations typical of rich open clusters and 
moderate-size globular clusters. 
In each case we find that the fraction of binaries in the core of a cluster does 
not decrease as the cluster evolves. 
In fact the overriding trend is for an increase in core binary fraction 
from the primordial value. 
Thus we do not agree with Ivanova et al. (2005) that the binary fraction in the core will 
be depleted in time. 
We also do not agree that models of globular cluster evolution 
need necessarily include large populations of primordial binaries. 

Our simulations have shown that the binary population in the core of a 
cluster is continually being replenished by stars from outside the core, 
many of which were previously in the core. 
This is a process we have termed {\it binary convection}. 
We also find that the binary content of an evolved star cluster is dominated 
by exchange binaries provided that the stellar density is relatively high. 
This is true of our moderate-size globular cluster models and we expect 
it to be true in more massive clusters. 
We also show that increasing the primordial binary fraction does not 
necessarily lead to an increase in the final binary fraction -- in fact it 
gives more scope for binary depletion. 
A key and paradoxical result is that a final binary fraction that can be achieved by 
choosing a higher primordial binary fraction may also be replicated 
by choosing an initially lower binary fraction. 

We find that the overall binary fraction of a cluster does not vary appreciably from 
the primordial value as a cluster evolves. 
This is a result of binary destruction being balanced by a greater rate of 
escape of single stars compared to binaries. 
We also find that the primordial binary frequency of a cluster is well preserved 
outside of the cluster half-mass radius. 
Therefore, observations of the current binary fraction in these regions is a 
good indicator of the primordial binary fraction while determination of the 
core binary fraction provides an upper limit. 

\acknowledgments

We acknowledge the generous support of the Cordelia Corporation and
that of Edward Norton which has enabled AMNH to purchase  
GRAPE-6 boards and supporting hardware. 
We thank the anonymous referee for extremely helpful comments 
and especially for alerting us to the scaling considerations.

\clearpage

\clearpage

\begin{deluxetable}{rrlrlcrl}
\tablecolumns{8}
\tablewidth{0pc}
\tablecaption{
Details of the four $N$-body simulations utilised in this work. 
Columns 1 and 2 show the number of single stars and binaries in the 
starting model. 
The distribution used to select the orbital separations of the primordial binaries 
is given in column~3 and this is followed by the maximum applied to the 
distribution (in au). 
Column~5 lists the primordial binary fraction and 
in column~6 we show the typical stellar density in the core for the simulation 
(stars/${\rm pc}^3$). 
The half-life of the simulation (time in Myr for $N_{\rm s} + N_{\rm b}$ to drop 
to half the initial value) is given in column~7 and finally an identifying label 
is supplied for each simulation. 
\label{t:table1}
}
\tabletypesize\normalsize
\tablehead{
 $N_{{\rm s},0}$ & $N_{{\rm b},0}$ & $\psi (a)$ & $a_{\rm max}$ & 
 $f_{{\rm b},0}$ & $n_{\rm c}$ & $t_{1/2}$ & label  
}
\startdata
95000 & 5000 & EFT30 & 100 & 0.05 & $10^2 - 10^4$ & $8920$ &  K100-5 \\ 
90000 & 10000 & EFT30 & 100 & 0.10 & $100 - 500$ & $8850$ & K100-10 \\  
40000 & 10000 & EFT30 & 50 & 0.20 & $10^3$ & $5560$ & K50-20 \\ 
12000 & 12000 & $\log a$ & 50 & 0.50 & $100 - 350$ & $2060$ & K24-50 \\
\enddata
\end{deluxetable}

\end{document}